\documentclass[draft,numberedheadings]{aipproc}

\pdfoutput=1

\newcommand{\bi}{\begin{itemize}}
\newcommand{\ei}{\end{itemize}}
\newcommand{\be}{\begin{equation}}
\newcommand{\ee}{\end{equation}}

\newcommand{\ben}{\begin{eqnarray}}
\newcommand{\een}{\end{eqnarray}}
\newcommand{\benstar}{\begin{eqnarray*}}
\newcommand{\eenstar}{\end{eqnarray*}}

\def\dbar{{\mathchar'26\mkern-12mu d}}

\layoutstyle{6x9}

\begin{document}

\title{Recent progress in fluctuation theorems and free energy recovery}


\author{A. Alemany}{
  address={Small Biosystems Lab, Departament de Fisica Fonamental, Facultat de Fisica, Universitat de Barcelona, Diagonal 647, 08028 Barcelona, Spain}
}

\author{M. Ribezzi}{
  address={Small Biosystems Lab, Departament de Fisica Fonamental, Facultat de Fisica, Universitat de Barcelona, Diagonal 647, 08028 Barcelona, Spain}
}

\author{F. Ritort}{
  address={Small Biosystems Lab, Departament de Fisica Fonamental, Facultat de Fisica, Universitat de Barcelona, Diagonal 647, 08028 Barcelona, Spain},
email={fritort@gmail.com},
altaddress={CIBER-BBN, Networking Center of Bioengineering, Biomaterials and Nanomedicine, ISCIII, Madrid, Spain},
}

\begin{abstract}
In this note we review recent progress about fluctuation relations and their applicability to free energy recovery in single molecule experiments. We underline the importance
of the operational definition for the mechanical work and the
non-invariance of fluctuation relations under
Galilean transformations, both aspects currently amenable to experimental
test. Finally we describe a generalization of the Crooks fluctuation
relation useful to recover free energies of partially equilibrated states and
thermodynamic branches. 
\end{abstract}

\maketitle

\section{Nonequilibrium small systems}
\label{NES}
In 1944 Erwin Schr\"odinger published the classic monography {\it What
  is life?} where he pointed out the importance of physical and chemistry laws to
understand living systems \cite{Sch44}. The notion that genetic information should
be encoded in an ``aperiodic crystal'' seeded the subsequent discovery of the
double helix structure of DNA. Chapter 7 of Schr\"odinger's  monography contains an
interesting discussion about the similarities and differences between
a clockwork motion and the functioning of an organism. According to
Schr\"odinger the regular
motion of a clock must be secured by a weak spring. Yet, whatever the
weakness of the spring is, it will produce frictional effects that do compensate
for the external driving of the clock (e.g. the battery) in order to secure its regular
motion. Being friction a statistical phenomenon he concludes that the
regular motion of the clock cannot be understood without statistical
mechanics. Then he further states: {\it For it must not believed that
the driving mechanism really does away with the statistical nature of
the process. The true physical picture includes the possibility that
even a regularly going clock should all at once invert its motion and,
working backward, rewind its own spring -at the expense of the heat of
the environment. The event is just `still a little less likely' than a
`Brownian fit' of a clock without driving mechanism.}

Recent advances in microfabrication techniques, detection systems and
instrumentation have made possible the measurement of such ``inverted
motions'' referred by Schr\"odinger. The controlled manipulation and detection of
very small objects makes possible to exert and measure tiny forces
applied on them and follow their trajectories in
space-time with resolution of piconewtons, nanometers and microseconds respectively. According to
the equipartition law, systems with a low number of degrees of freedom
embedded in a thermal environment exhibit energy fluctuations that are a
few times $k_{\rm B}T$ ($k_{\rm B}$ being the Boltzmann constant and $T$ the
environmental temperature).  Techniques such as atomic
force microscopy (AFM), optical tweezers (OT) and magnetic tweezers (MT) allow for the
controlled manipulation of individual molecules such as nucleic acid
structures and proteins \cite{Rit06}, the measurement of very small energies (within the
$k_{\rm B}T$ scale) \cite{HugBizForSmiBusRit10} and the observation of ``inverted motion'' in
translocating enzymes \cite{BusLipRit05}.  These developments during the past years have
been accompanied by a concomitant progress of theoretical results in the
domain of nonequilibrium physics \cite{Rit08}. This
contribution reviews some of the most basic concepts around fluctuation
theorems and their experimental verification.

\subsection{Control parameters, configurational variables and the definition of work}
\label{controlwork}
\begin{figure}
\includegraphics[scale=0.51]{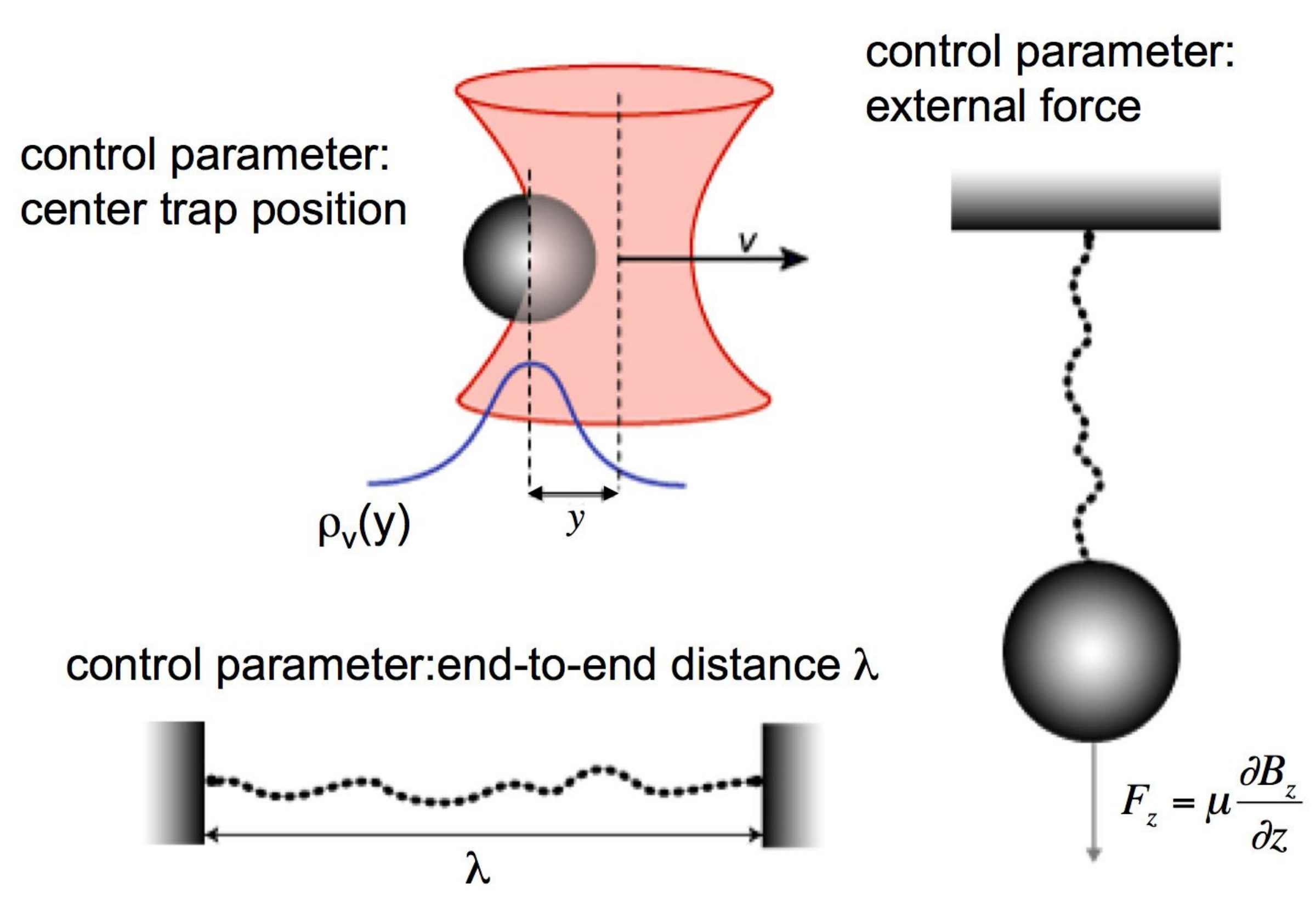}
\caption{{\bf Control parameter and configurational variables.}
  Different experimental setups corresponding to different types of
  control parameters (denoted as $\lambda$) or configurational variables (denoted as $x$). (Top left) A
  micron-sized bead dragged through water. $\lambda$ could
  be the center of the trap measured in the lab (i.e. fixed to the water) frame whereas $x$ is the displacement
  of the bead, indicated as $y$, respect to the center of the trap. (Bottom left) A polymer tethered between
  two surfaces. $\lambda$ is the distance between the
  surfaces and $x$ the force acting on the polymer. (Right) A polymer stretched with magnetic tweezers. $\lambda$ is the force acting on the magnetic bead and $x$ is the molecular extension of the tether. Figure taken from \cite{BusLipRit05}.}
\label{fig1}
\end{figure}

In small systems it is crucial to make a distinction between controlled parameters
and non-controlled or fluctuating variables. Controlled parameters are
those macroscopic variables that are imposed on the system by the
external sources (e.g. the thermal environment)
and do not fluctuate with time. In contrast, non-controlled
variables are microscopic quantities describing the internal
configuration of the system and do fluctuate in time because they are
subject to Brownian forces. Let us consider a typical single molecule
experiment where a protein is pulled by an AFM. In this case the control parameter is
given by the position of the cantilever that determines the degree of
stretching and the average tension applied to the ends of the
protein. Also the temperature and the pressure inside the fluidic
chamber are controlled parameters. However, the height of the tip
respect to the substrate or the force acting on the protein are fluctuating variables
describing  the molecular extension of
the protein tethered between tip and substrate. Also the position of each
of the residues along the polypeptide chain are fluctuating
variables. Both molecular extension or force and the residues positions define
different types of configurational variables. However only the former are subject
to experimental measurement and therefore we will restrict our
discussion throughout this paper to such kind of experimentally accessible
configurational variables. Figure \ref{fig1} illustrates other examples of
control parameters and configurational variables. In what follows we
will denote by $\lambda$ the set of controlled (i.e. non-fluctuating)
parameters and $x$ the set of configurational (i.e. fluctuating) variables. The
definition of what are controlled parameters or configurational variables is
broad. For example, a force can be a configurational variable and a
molecular extension can be a controlled parameter, or vice versa,
depending on the experimental setup (see Figure \ref{fig1}, right example).

The energy of a system acted by external sources can be
generally described by a Hamiltonian or energy function, $U(x,\lambda)$. The net
variation of $U$ is given by the conservation law,
\be
dU=\left(\frac{\partial U}{\partial x} \right)dx+\left(\frac{\partial
  U}{\partial \lambda}
\right)d\lambda=\dbar Q+\dbar W
\label{eqQW}
\ee
where\phantom{a} $\dbar Q$,\phantom{a} $\dbar W$ stand for the infinitesimal heat and work
transferred to the system.  The previous mathematical relation has
simple physical interpretation. Heat accounts for the energy transferred to the
system when the configurational variables change at fixed value of the control
parameter. Work is the energy delivered to the system by the external
sources upon changing the control parameter for a given
configuration. The total work performed by the sources on the system when the
control parameter is varied from $\lambda_i$ to $\lambda_f$ is given by,
\be
W=\int_{\lambda_i}^{\lambda_f}\dbar W=\int_{\lambda_i}^{\lambda_f}\left(\frac{\partial
  U}{\partial \lambda}\right)d\lambda=\int_{\lambda_i}^{\lambda_f}F(x,\lambda)d\lambda
\label{eqtotW}
\ee
where $F(x,\lambda)$ is a generalized force defined as,
\be
F(x,\lambda)=\left(\frac{\partial
  U}{\partial \lambda}\right)~~~~~~.
\label{eqforce}
\ee
It is important to stress that the generalized force is not
necessarily equal to the mechanical force acting on the system. In other
words, $F(x,\lambda)$ is a configurational dependent variable conjugated
to the control parameter $\lambda$ and has dimensions of
[energy]/[$\lambda$] which are not necessarily Newtons. In the
example shown in the right of Figure \ref{fig1} the control parameter is the magnetic force
$\lambda\equiv f$ and the configurational variable $x$ is the molecular
extension of the polymer. The total Hamiltonian of the system is then given by
$U(x,f)=U_0(x)-fx$ where $U_0(x)$ is the energy of the system at
$\lambda=f=0$. In other words, the external force $f$ shifts all
energy levels (defined by $x$) of the original system by the amount
-$fx$. The generalized force is then given by $F(x,f)=-x$ (i.e. it has
the dimensions of a length) and\phantom{a} $\dbar W=-xdf$. The fact that\phantom{a} $\dbar
W$ is equal to $-xdf$ and not equal to $fdx$ has generated some
controversy \cite{Pel08}. Below we show how this distinction is already important for
the simplest case of a bead in the optical trap. In Section \ref{acctrans} we also
show how the physically sound definition of mechanical work is amenable
to experimental test.

\subsection{A classical experiment: the bead in the optical trap}
\label{beadintrap}
\begin{figure}
\includegraphics[scale=0.4]{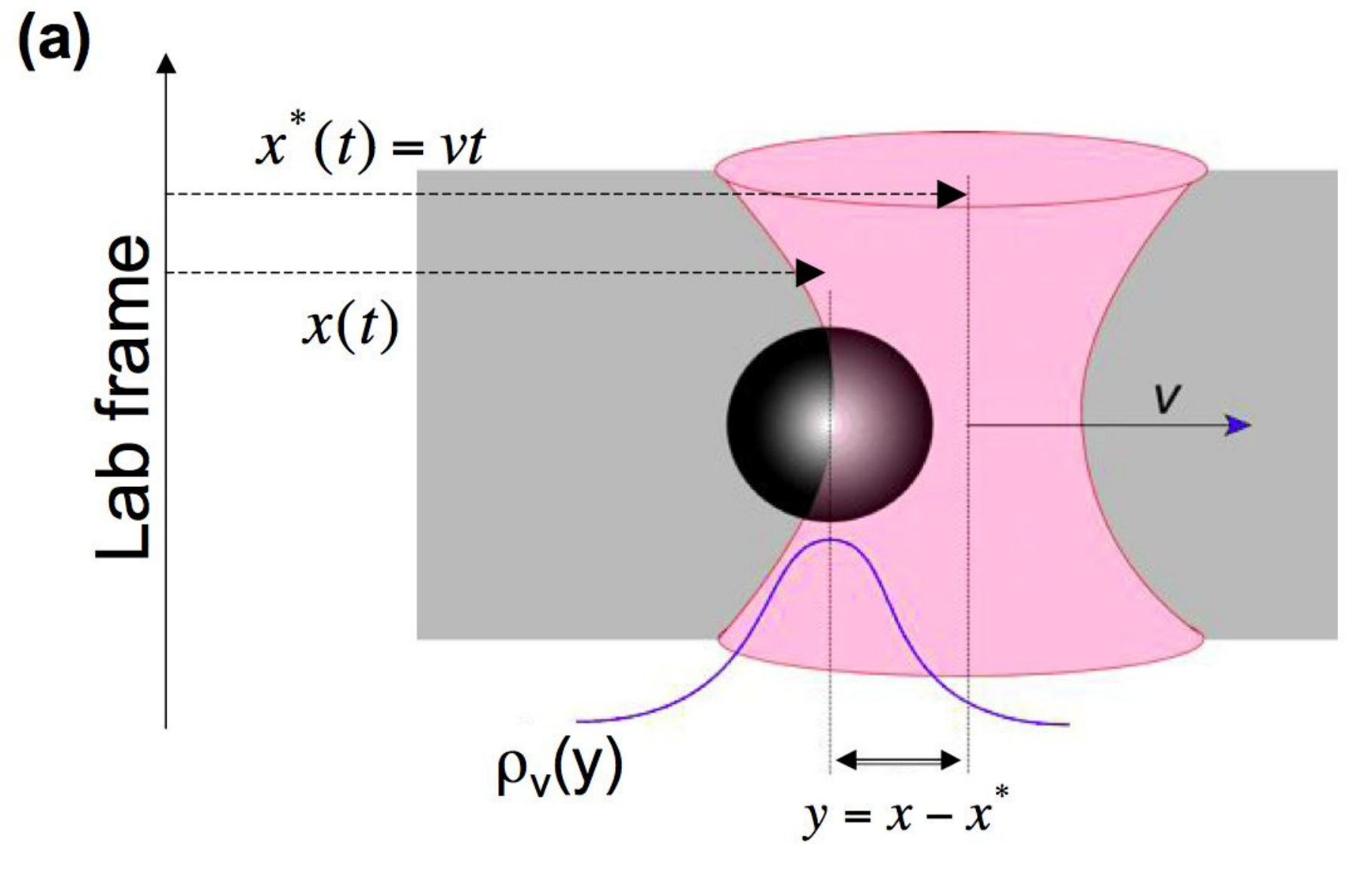}\includegraphics[scale=0.3]{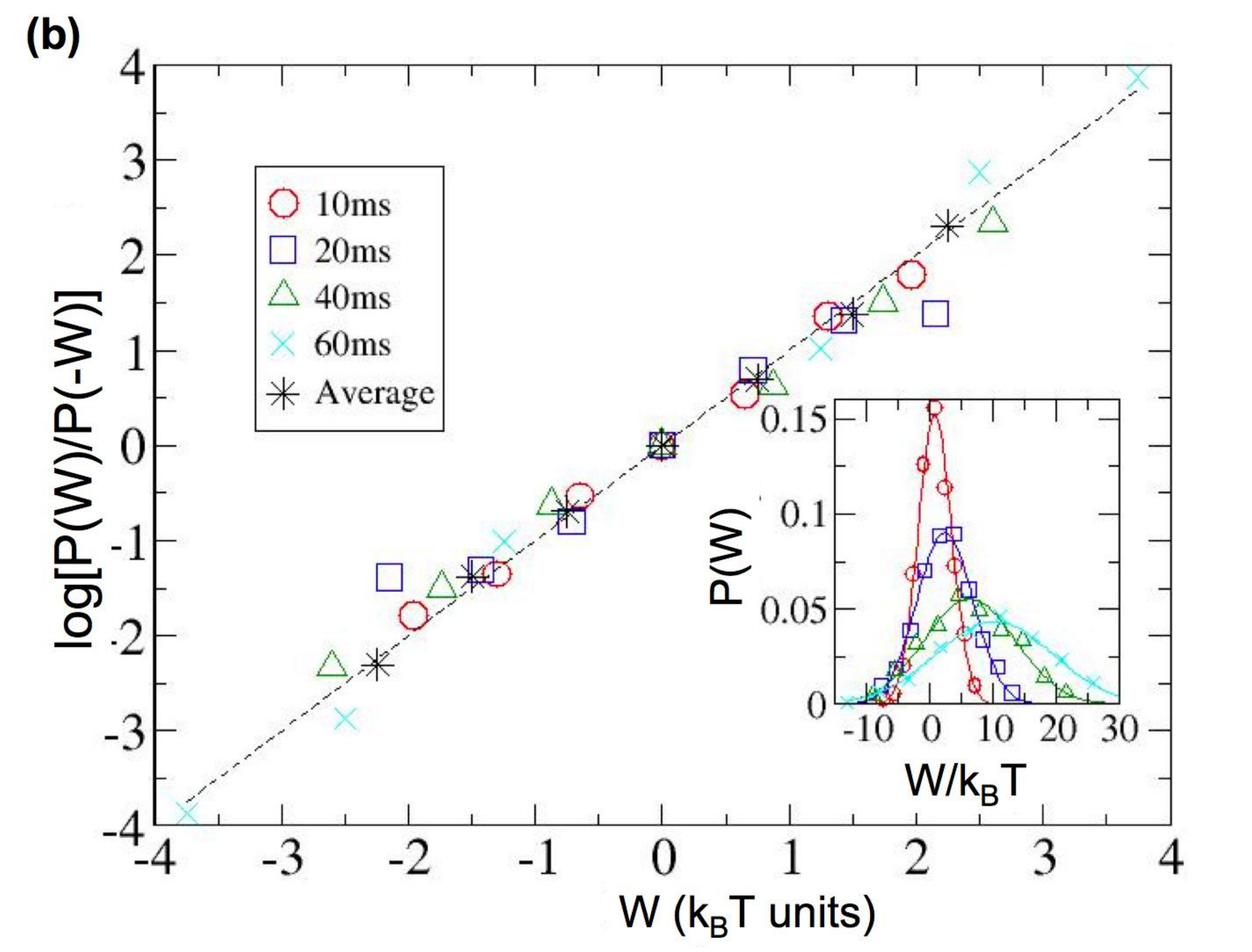}
\caption{{\bf The bead in the optical trap dragged through water. }(a) Variables defining the experiment. (b) Work distribution measurements corresponding to different elapsed times. The mean work $\overline{W}$ and variance $\sigma_W^2$ increase with time (Inset). The main panel shows the experimental test of the fluctuation relation Eq.(\ref{eqfr}) for all experimental data put together.}
\label{fig2}
\end{figure}

In 2002 Dennis Evans and coworkers in Camberra (Australia) performed
the first experiment where the ``inverted motions'' could be observed \cite{Wan02}. The
experiment is shown in Figure \ref{fig2}a. A micron-sized spherical bead made of silica or
polystyrene is immersed in water inside a fluidic
chamber and captured in an optical trap of infrared light generated by a high numerical
aperture objective. Initially the trap is at a rest position and the
bead is in thermal equilibrium and fluctuating around the center trap position. Suddenly the trap is set into motion at
a constant speed $v$ and the bead is dragged through the water. After a
transient time $\tau=\gamma/k$ the bead reaches a steady state where the
Stokes frictional force is counter balanced by the optical trapping
force. The average bead position lags behind the center of the trap by a distance
$\overline{y}=\gamma v/k$ where $\gamma=6\pi\eta R$ is the friction coefficient
($\eta$ is the water viscosity and $R$ is the bead radius) and $k$ is
the stiffness of the trap. In the laboratory frame (see Figure \ref{fig2}a) the
bead and trap center have coordinates $x(t)$ and $x^*(t)=vt$ (we take
$x=x^*=0$ at $t=0$ when the trap is set in motion). The distance between
the center of the trap and the bead is $y=x^*-x$ and the restoring force acting on
the bead is given by $F(y)=ky$. In this example the control parameter is
given by the trap center $\lambda=x^*$ whereas $x$ is the
configurational variable.  The trapping energy of the bead
is given by $U(x,\lambda)=(1/2)k(x-\lambda)^2$ and  the generalized force
$F=k(\lambda-x)=ky$ (cf. Eq.(\ref{eqforce})). The work exerted by the
trap on the bead is then equal to $W=\int_0^tF(y)dx^*=vk\int_0^ty(s)ds$. The
first remarkable fact in this expression is that the work $W$ is neither
equal to $W'=\int_0^tF(y)dx$ nor $W''=\int_0^tF(y)dy$. These three quantities have different physical meaning. 
In fact, by exactly integrating the force, $W''$ becomes
equal to the energy difference between the initial and final
configurations. Whereas $W'$ is equal to minus the heat, $-Q$. Since
$dy=dx^*-dx$ what we are now facing is again the mathematical statement
of energy conservation. Note
also that the work definition is non-invariant under Galilean
transformations. In fact, the work
definition  requires that $x$, as measured in the lab frame, is the proper
configurational variable. If we choose $y$ rather than $x$ ($y$ is now
measured in the trap-moving frame) then
$U(y,\lambda)=(1/2)ky^2$ is independent of $\lambda$ and the work would
be identically zero which makes no sense.

By repeating the moving trap experiment an infinite number of times a work
distribution will be produced. The shape of the work distribution must
be Gaussian because the stochastic variable $y$ follows an Ornstein-Uhlenbeck process
that can be described by a linear Langevin equation. Consequently, only the first and
second cumulants of the work distribution are non-zero. Let us note in
passing that the Gaussian property is not fulfilled by the heat and the energy difference,
which are known to exhibit exponential tails \cite{ZonCoh03,GarCil05}. The mean work $\overline{W}$ and variance
$\sigma_W^2=\overline{W^2}-\overline{W}^2$ can be easily worked out in the
asymptotic regime for times $t$ that are longer than the relaxation time of the bead, $t>>\gamma/k$. In
this limit, 
\begin{eqnarray}
\overline{W}=vk\int_0^t\overline{y(s)}ds\to vk\overline{y}t=\gamma v^2t\label{eqw1}\\
\sigma_W^2=\overline{W^2}-\overline{W}^2=v^2k^2\int_0^t\int_0^t\overline{y(s_1)y(s_2)}ds_1ds_2\to
2v^2k^2t\int_0^t\overline{y(0)y(s)}ds=\nonumber\\
=2v^2k^2t\overline{y^2}\tau=2v^2k^2t\frac{k_{\rm B}T}{k}\frac{\gamma}{k}=2k_{\rm B}T\gamma
v^2t
\label{eqw2}
\end{eqnarray}
where $\overline{...}$ stands for an average over trajectories. In deriving
Eq.(\ref{eqw2}) we used time-translational invariance in the steady
state and the result 
$\overline{y(0)y(s)}=\overline{y(0)^2}\exp(-s/\tau)$ in the steady state 
($\tau=\gamma/k$ being the bead relaxation time) with
$\overline{y(0)^2}=k_{\rm B}T/k$ due to the equipartition law. The work probability
distribution is finally given by, 
\be
P(W)=\frac{1}{\sqrt{2\pi \sigma_W^2}}\exp\left( -\frac{(W-\overline{W})^2}{2\sigma_W^2} \right)~~~.
\label{eqpw}
\ee
These relations teach us
various things:
\begin{enumerate}

\item{\bf Second law.} The mean work $\overline{W}$ is always positive (second law) and
only vanishes at all times for
$v\to 0$, i.e. when the trap is moved in a quasistatic way. 

\item{\bf Observation of ``Inverted motions''.} Although $\overline{W}>0$ there are always
  trajectories for which $W<0$. These are the ``inverted
  motions'' refereed to by Schr\"odinger and recently renamed as
  ``transient violations of the second law''. Along these ``inverted
  motion'' trajectories the bead extracts energy from  heat fluctuations to overcome the
  frictional forces and to move ahead
  of the center of the trap.    

\item{\bf ``Inverted motions'' as rare events.} Both the mean work and the standard deviation of the work increase with time
and trap speed. However the standard deviation $\sigma_W$ increases as $\sqrt{t}$ whereas
$\overline{W}$ increases faster (linearly with $t$). Therefore, although
it is possible to find trajectories where $W<0$, these are rare events
because their relative fraction
decreases exponentially fast with time. $W<0$ trajectories become more
probable (i.e. less rare) at short
times. In the limit $t\to 0$ they reach the 50\% of all events.

\item{\bf Fluctuation relation.} The work probability density function shown in Eq.(\ref{eqpw})
  satisfies a fluctuation relation. From Eqs.(\ref{eqw1},\ref{eqw2}) we find $\sigma_W^2=2k_{\rm B}T
  \overline{W}$. It is straightforward to check that the following relation holds,
\be
\frac{P(W)}{P(-W)}=\exp\left(\frac{W}{k_{\rm B}T}\right)~~~~.
\label{eqfr}
\ee
Eq.(\ref{eqfr}) receives the name of a fluctuation relation because it is an exact
mathematical relation describing arbitrarily large work
fluctuations. Eq.(\ref{eqfr}) was derived from Eq.(\ref{eqpw}) which was
obtained in the
limit of long enough times. More elaborate calculations show that this
relation is exact for arbitrary times \cite{MazJar99,ZonCoh03}. 

\end{enumerate}

In Figure \ref{fig2}b we show an experimental test of these results. The
fluctuation relation in Eq.(\ref{eqfr}) corresponds to a special case of
what is known as
transient fluctuation theorem (TFT) \cite{EvaSea94}. The system is initially in
equilibrium and transiently driven out of equilibrium by external
forces. The generalization of such relation to include arbitrary
nonequilibrium transient states gives the fluctuation relation by Crooks
described in the next section.

\section{The Crooks fluctuation relation and free energy recovery.}
\label{cfr}
\begin{figure}
\includegraphics[scale=0.6]{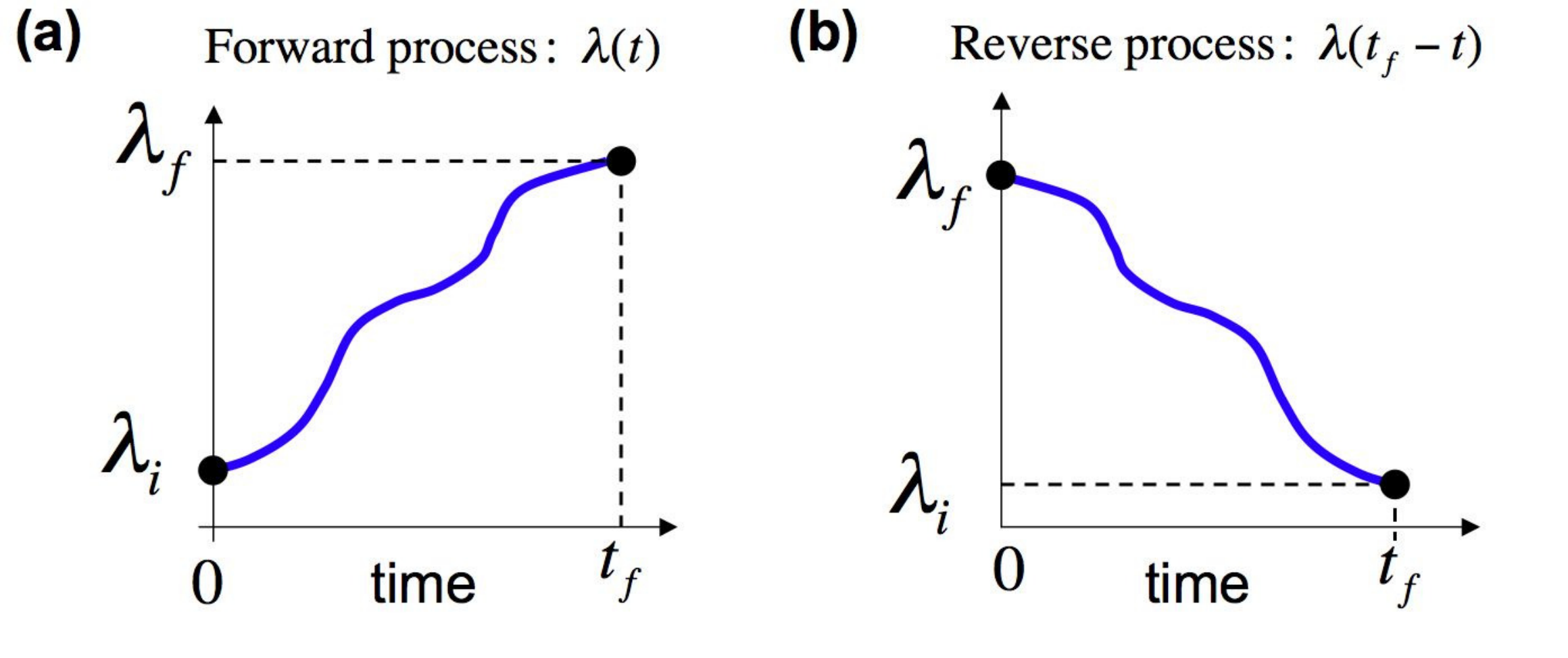}
\caption{{\bf Forward and reverse paths. }(a) An arbitrary forward protocol. The system starts in equilibrium at $\lambda_i$ and is transiently driven out of equilibrium until $\lambda_f$. At $\lambda_f$ the system may be or not in equilibrium. (b) The reverse protocol of (a).  The system starts in equilibrium at $\lambda_f$ and is transiently driven out of equilibrium until $\lambda_i$. At $\lambda_i$ the system may be or not in equilibrium.}
\label{fig3}
\end{figure}

Let us consider a generic system in thermal equilibrium that is
transiently driven out of equilibrium during the time interval $[0,t_f]$ by varying $\lambda$
according to a protocol $\lambda(t)$ from an initial value
$\lambda(0)=\lambda_i$ to a final value $\lambda(t_f)=\lambda_f$. We
refer to this as the forward
(F) process. By repeating this process an infinite number of times we generate the work
distribution $P_F(W)$. Let us consider now the reverse process where the system
starts in equilibrium at $\lambda_f$ and $\lambda$ is varied
according to the time reversal protocol, $\lambda(t_f-t)$, until reaching
the final value $\lambda_i$ (see Figure \ref{fig3}). The
reverse (R) process can be repeated an infinite number of times to produce the work
distribution $P_R(W)$. The Crooks fluctuation relation (CFR) reads \cite{Crooks99}, 
\be
\frac{P_F(W)}{P_R(-W)}=\exp\left(\frac{W-\Delta G}{k_{\rm B}T}\right)
\label{eqcrooks}
\ee
where $\Delta G=G(\lambda_f)-G(\lambda_i)$ is equal to the free energy
difference between the equilibrium states at $\lambda_f$ and
$\lambda_i$. Eq.(\ref{eqfr}) is a particular case of the CFR where
$P_F(W)=P_R(W)$ (the trapping potential is symmetric $V(y)=V(-y)$) and $\Delta G=0$ (the free energy of the bead in the
trap does not depend on the position of the trap, $x^*$). A particular
result of the CFR is the well-known Jarzynski equality \cite{Jar97},
$\overline{\exp(-W/k_{\rm B}T)}=\exp(-\Delta G/k_{\rm B}T)$, that has been used for
free energy recovery \cite{HumSza01,LipDumSmiTinBus02} by inverting the mathematical identity, $\Delta
G=-k_{\rm B}T\log(\overline{\exp(-W/k_{\rm B}T)})$. However this expression is
strongly biased for a finite number of measurements \cite{ZucWoo02,RitBusTin02}. Bidirectional
methods that combine information from the forward and reverse processes
and use the CFR have proven more predictive \cite{Ben76,ShiBaiHooPan03,MinAdi08}. In particular the CFR immediately
implies that $P_F(W)=P_R(-W)$ for $W=\Delta G$ showing that it is
possible to measure $\Delta G$ in irreversible processes by measuring
the forward and reverse irreversible work distributions and looking for
the value of $W$ where they cross each other. The CFR was experimentally
tested in 2005 in RNA pulling experiments with laser tweezers \cite{ColRitJarSmiTinBus05} showing
this to be a reliable and useful methodology to extract free energy
differences between states that could not be measured with bulk methods.

\begin{figure}
\includegraphics[scale=0.6]{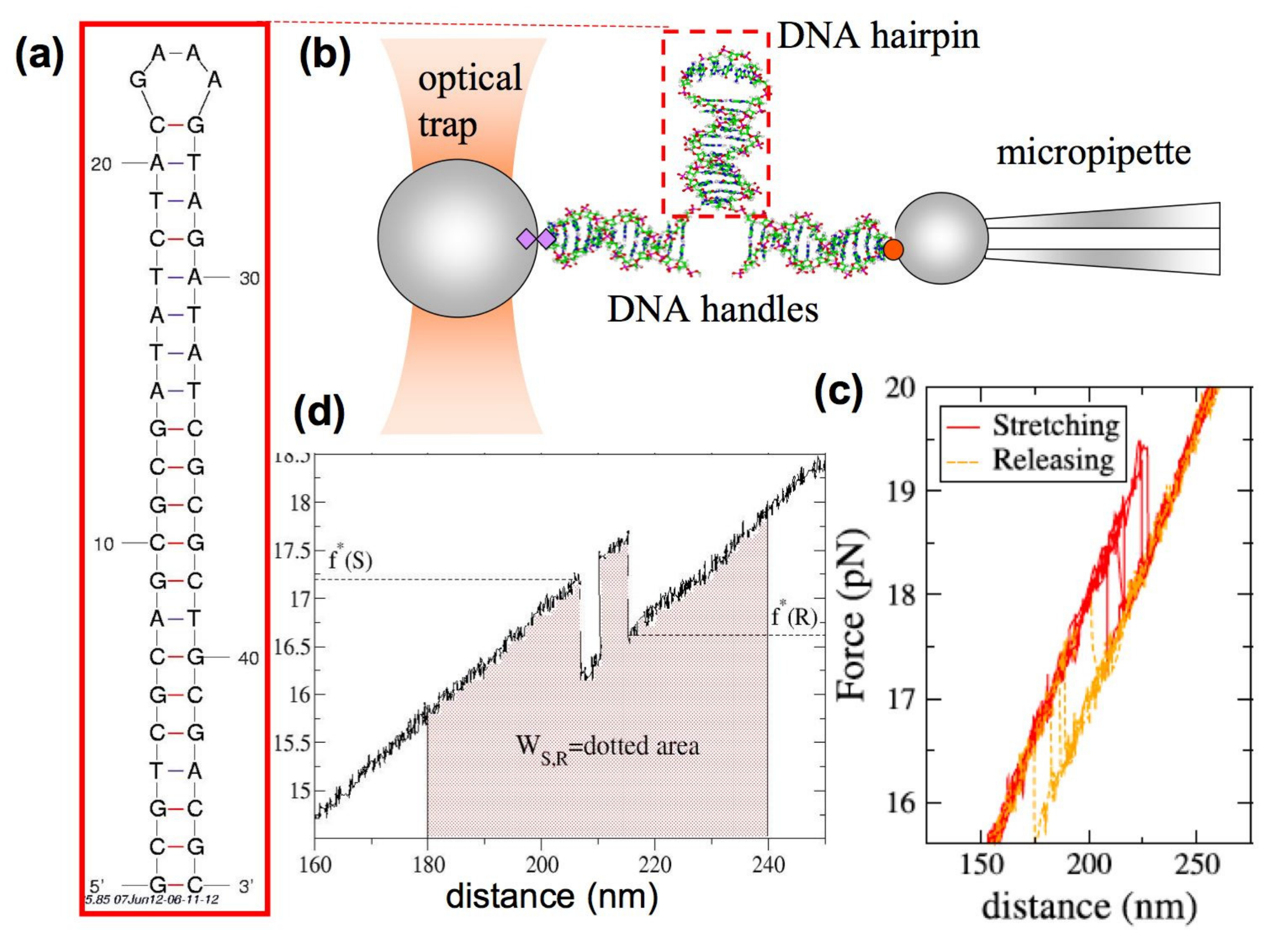}
\caption{{\bf Mechanical folding/unfolding of DNA hairpins.}(a) The
  sequence of a DNA hairpin with a 21bp stem ending in a tetraloop. (b)
  Experimental setup. A molecular construct made of the hairpin shown in (a)
  flanked by two dsDNA handles (29bp each) is tethered between two
  micron-sized beads. In the experiments the trap is moved relative to
  the pipette at speeds ranging from 10 to 1000nm/s. (c) Different force
  cycles recorded at 300nm/s. The red curves indicate the stretching
  parts of the cycle whereas the orange curves indicate the releasing
  parts of the cycle. Note that the forces of unfolding and refolding
  are random due to the stochastic nature of the thermally activated
  unfolding/folding process. The marked hysteresis is a signature of an
  irreversible process. (d) Measurement of work for a single
  trajectory. It is given by the area below the force-distance curve
  integrated between two trap positions. Trap distances are
  relative. Note that there might be more than one unfolding or
  refolding event along each trajectory. $f^*(S)(f^*(R))$ defines the first
  rupture force in the unfolding (refolding) process. Figure taken from \cite{MosManForHugRit09}.}
\label{fig4}
\end{figure}

\begin{figure}
\includegraphics[scale=0.35]{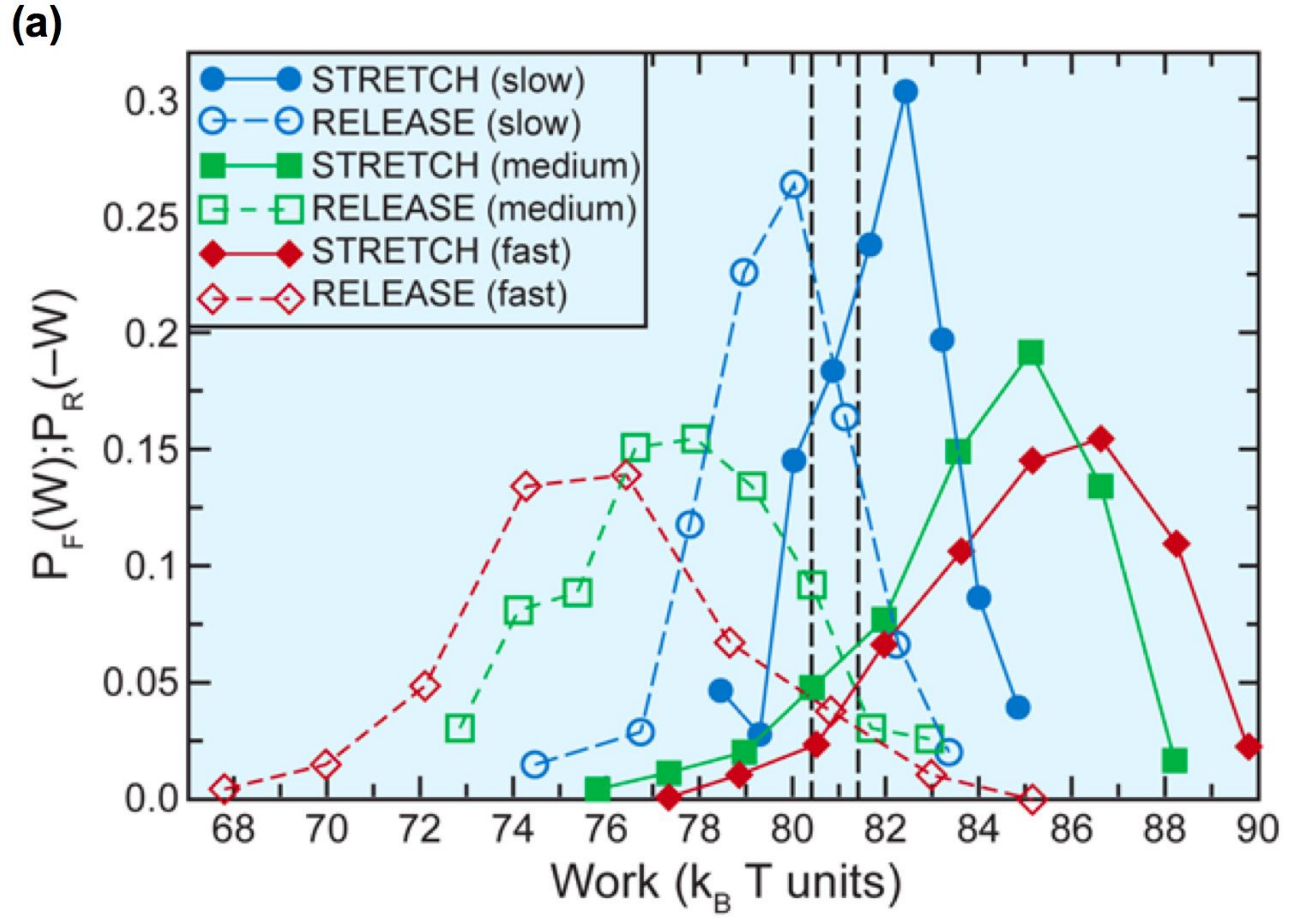}\includegraphics[scale=0.35]{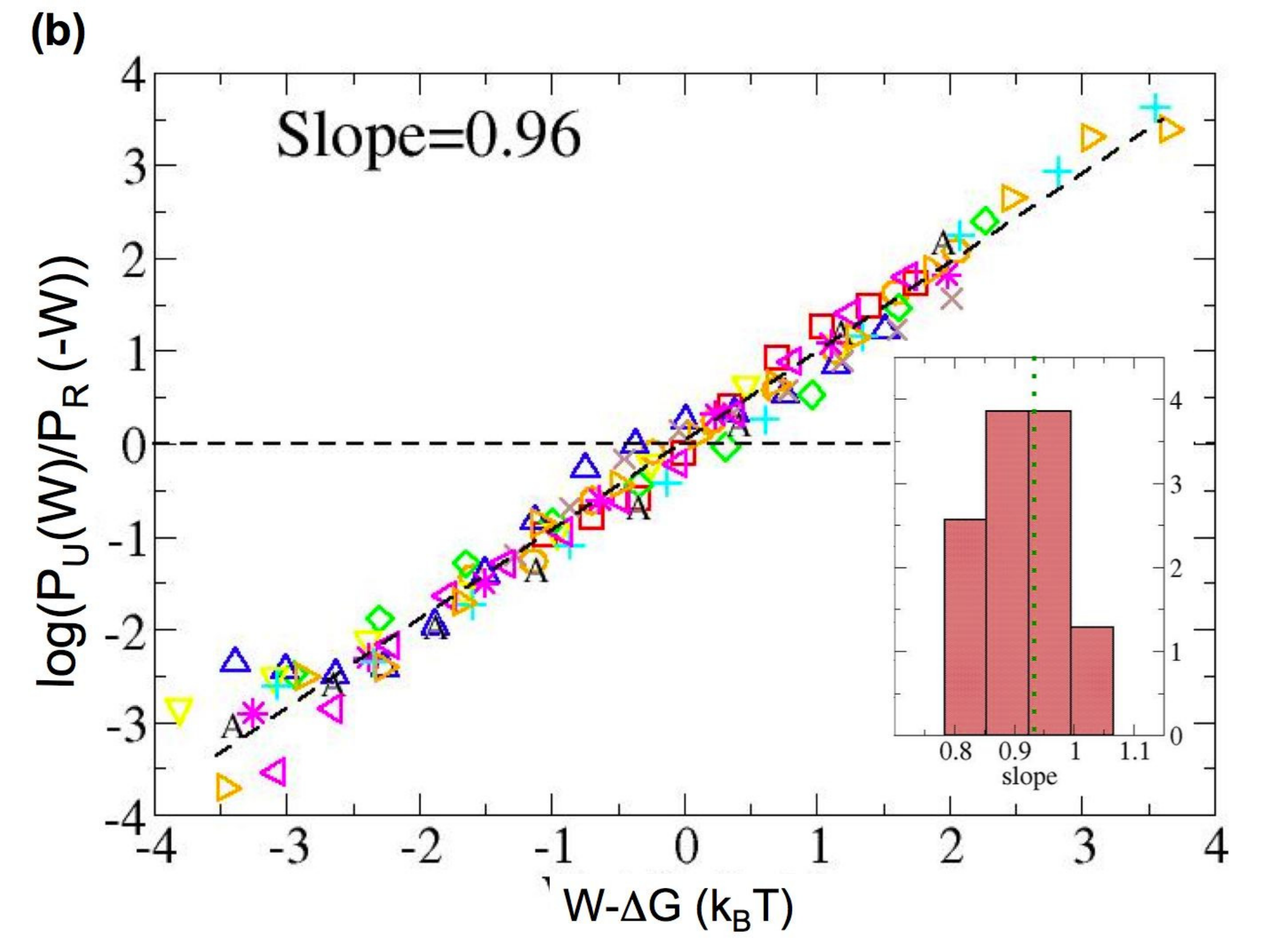}
\caption{{\bf The Crooks fluctuation relation.}(a) Work distributions
  for the hairpin shown in Figure \ref{fig4} measured at three
  different loading rates: 50 nm/s (blue), 100 nm/s (green) and 300 nm/s
  (red). Unfolding or forward (continuous lines) and refolding or
  reverse work distributions (dashed lines) cross each other at a a
  value $\simeq 81k_{\rm B}T$ independent of the loading rate.(b)
  Experimental test of the CFR for 10 different molecules pulled at
  different speeds. The log of the ratio between the unfolding and refolding work
  distributions is equal to $(W-\Delta G)$ in $k_{\rm B}T$ units. The inset
  shows the distribution of slopes for the different molecules which
  are clustered around an average value of 0.96. Figure taken from \cite{MosManForHugRit09}.}
\label{fig5}
\end{figure}

In Figure \ref{fig4} we summarize recent results obtained in the Small Biosystems lab in
Barcelona for the mechanical unfolding/refolding of DNA hairpins
\cite{MosManForHugRit09,ManMosForHugRit09} using a dual-beam
miniaturized optical tweezers \cite{HugBizForSmiBusRit10}. DNA
hairpins are versatile structures formed by a stem of a few tens of base
pairs that end in loop. They have some advantages as compared to RNA
structures such as the easier synthesis and larger chemical
stability. DNA hairpins can be easily synthesized and ligated to
dsDNA handles to produce a construct ready to be
pulled with the tweezers \cite{WooBehLarTraHerBlo06}.  By
chemically labeling the ends of the dsDNA handles it is possible to
tether a DNA construct (formed by the DNA hairpin inserted between the
two flanking handles) between two micron sized beads. One bead is
immobilized in the tip of a pipette. The other bead is captured in the
optical trap. The deflected light by the trapped bead provides a direct
measurement of the force applied on the molecule.  By repeatedly
steering the optical trap back and forth it is possible to unfold and
refold the hairpin structure many times until the tether breaks. The unfolding of the hairpin is revealed
by a sudden drop in the force due to the increase in molecular extension
from the released
single-stranded DNA of the hairpin. Such increase causes a retraction in the
position of the bead in the trap and a force drop. Analogously, when the hairpin refolds a sudden
increase in force is observed. One of
the most successful constructs we have designed in our lab consists of DNA
hairpins linked to two beads via extremely short (29bp) dsDNA handles \cite{ForLorManHugRit10}. These
constructs are found to moderately increase the signal-to-noise ratio of the
measurements allowing for precise work measurements. In a pulling experiment
the force versus the relative trap-pipette distance is recorded and the area
below that curve provides a direct measurement of the work. Repeated
measurements of the work make possible an experimental verification of
the CFR (see figure \ref{fig5}).

\subsection{About the right definition of work: accumulated versus
  transferred work}
\label{acctrans}
\begin{figure}
\includegraphics[scale=0.6]{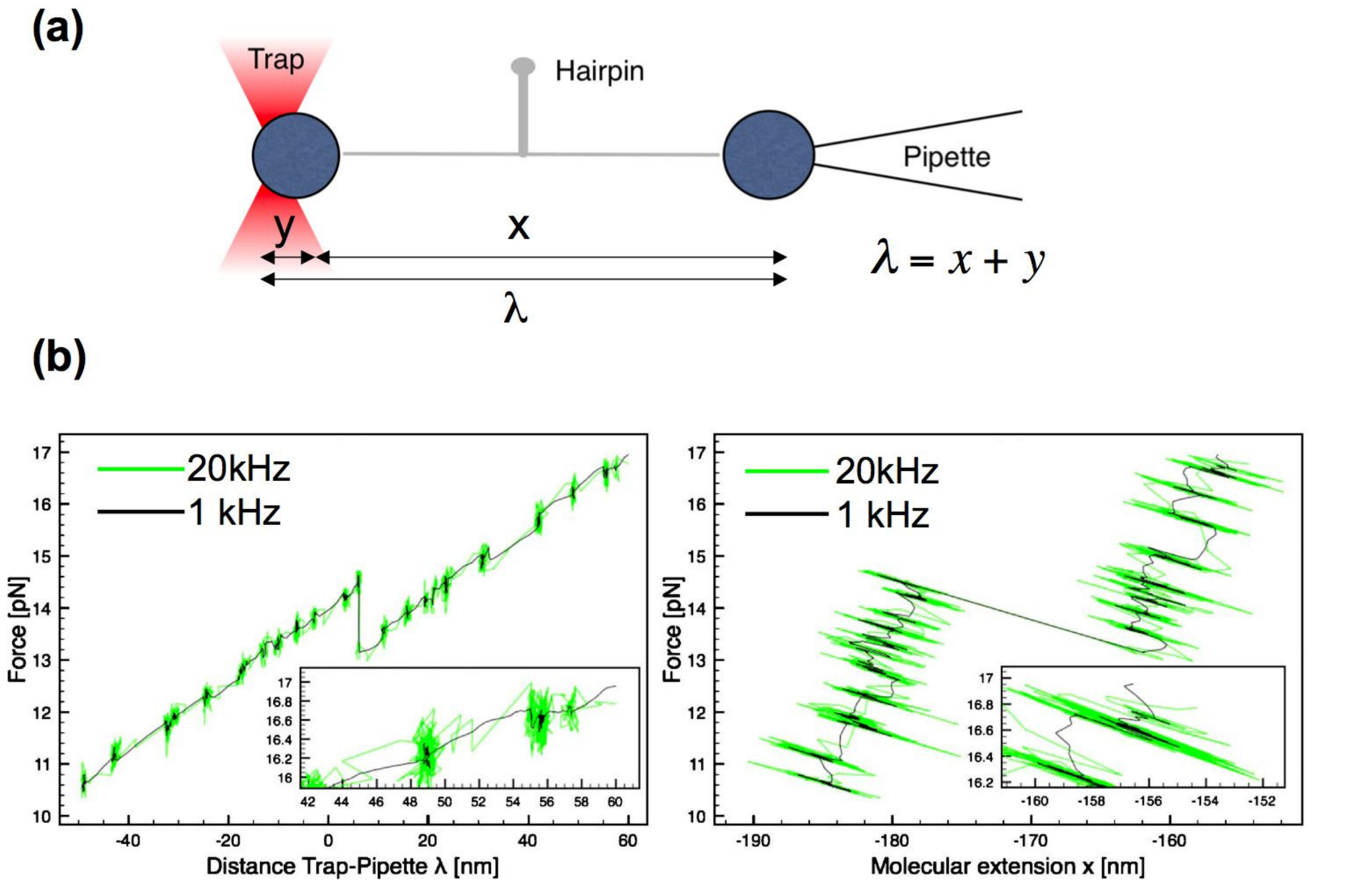}
\caption{{\bf FDC versus FEC. }(a) Experimental setup and different
  variables. (b) The FDC and FEC are defined as the curves obtained by
  plotting the force versus the trap position or the molecular
  extension respectively. Although force fluctuations in both types of
  curves show a dependence with the bandwidth of the measurement
  (black, 1kHz; green 20 kHz) only in the FEC the measurement of the
  work is very sensitive to such fluctuations. Figure taken from \cite{MosLorForHugRit09}.}
\label{fig6}
\end{figure}

In a pulling experiment there are two possible representations of the
pulling curves (Figure \ref{fig6}b). In one representation the force is
plotted versus the relative trap-pipette distance ($\lambda$), the so-called
force-distance curve (hereafter referred as FDC). In the other
representation the force is plotted versus the relative molecular
extension ($x$), the so-called force-extension curve (hereafter referred as
FEC). In the optical tweezers setup $\lambda=x+y$ where $y$ is the
distance between the bead and the center of the trap. The measured force
is given by $F=ky$ where $k$ is the stiffness of the trap. The
areas below the FDC and the FEC define two possible
work quantities, $W=\int_{\lambda_i}^{\lambda_f}
Fd\lambda$ and $W'=\int_{x_i}^{x_f}
Fdx$. From the relation $d\lambda=dx+dy$ we get,
\be
W=W'+W_b=W'+\frac{F_f^2-F_i^2}{2k}
\label{eqWW'}
\ee

\begin{figure}
\includegraphics[scale=0.6]{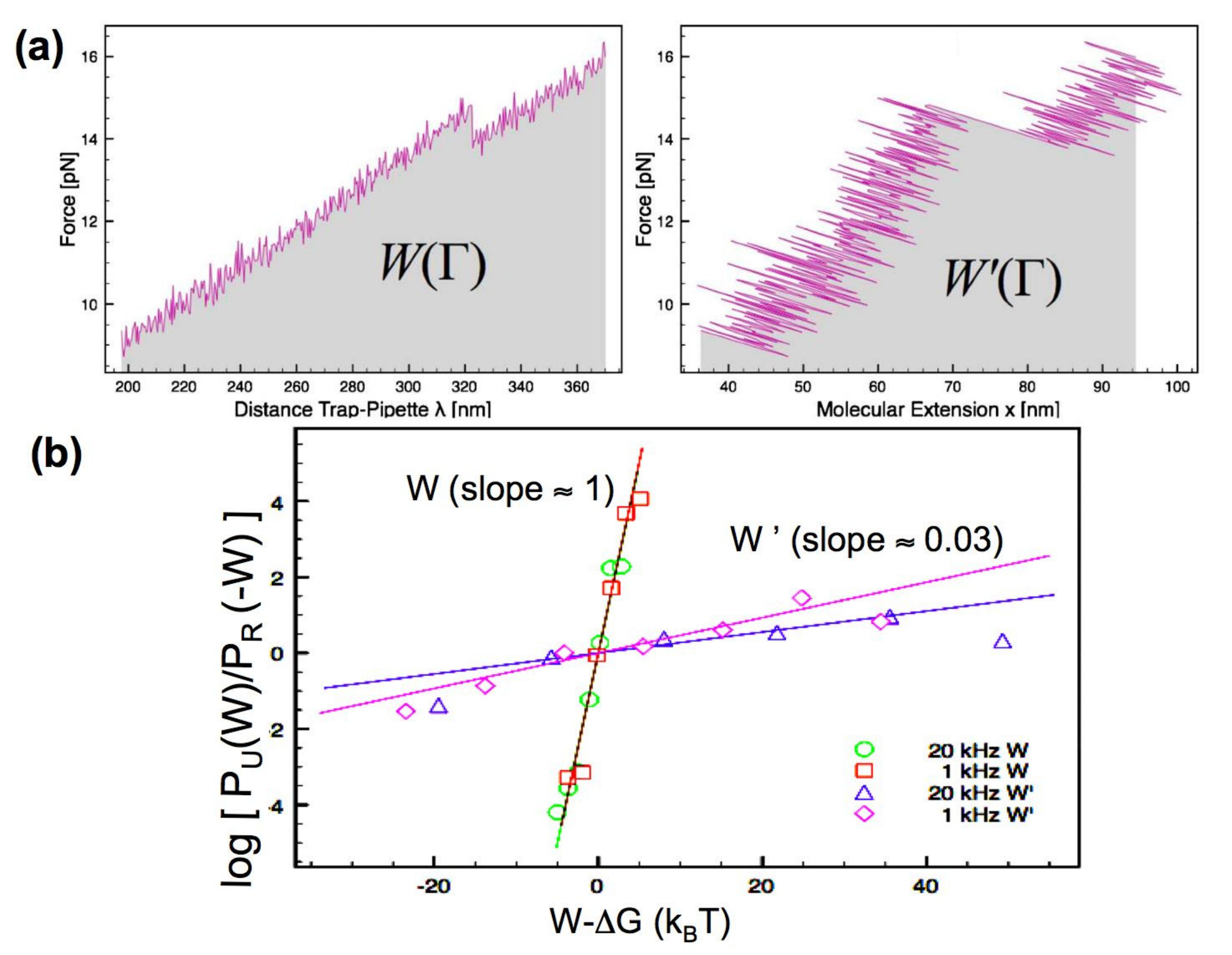}
\caption{{\bf Accumulated ($W$) versus transferred ($W'$) work.} (a)
  The two work quantities for a given experimental trajectory. Note
  that the effect of bandwidth dependent force fluctuations is much larger for $W'$
  as compared to $W$, showing the importance of the boundary term
  Eq.\ref{eqWW'}. (b) Experimental test of the CFR. When using $W$ the
  CFR is satisfied at all bandwidths. However when we use $W'$ the CFR
  is strongly violated and dependent on the measurement bandwidth. Figure taken from \cite{MosLorForHugRit09}.}
\label{fig7}
\end{figure}

where $F_i$, $F_f$ are the initial and final forces along a given
trajectory. $W$ is often called the total accumulated work and contains the work exerted to displace the bead in the trap, $W_b$, and the work
transferred to the molecular system, $W'$ (therefore receiving the  name
of transferred work) \cite{FujSch03,MosLorForHugRit09}. The term $W_b$ appearing in Eq.(\ref{eqWW'}) implies that $W$, $W'$ cannot simultaneously satisfy the
CFR. What is the right definition of the mechanical work? In other
words, which work definition satisfies the CFR? The problem we are
facing now is identical to the one we previously encountered in section
\ref{beadintrap} where we had to distinguish between work and heat. The
answer to our question is straightforward if we correctly identify which
are the control
parameters and which are the configurational variables. In the lab frame
defined by the pipette (or by the fluidic chamber to which the
pipette is glued) the control parameter $\lambda$ is given by the
relative trap-pipette distance, whereas the molecular extension $x$ stands for the
configurational variable. Note that, due to the non-invariance
property of the CFR under Galilean transformations, $y$ cannot be used as
configurational variable because it is defined respect to the co-moving
frame defined by the trap. The same problem was found in section
\ref{beadintrap} when comparing the distances $x$ and $y$ for the bead
in the trap. The total energy of the molecular system is then given by
$U(x,\lambda)=U_m(x)+(k/2)(\lambda-x)^2$ where $U_m(x)$ is the energy of
the molecular system. From Eq.(\ref{eqforce}) and
using $\lambda=x+y$ we get $F=ky=k(\lambda-x)$. From
Eq.(\ref{eqtotW}) we then conclude that the mechanical work that
satisfies the CFR is the accumulated work $W$ rather than the
transferred work $W'$. We remark a few relevant facts,

\begin{enumerate}

\item{\bf The transferred work $W'$ does not satisfy the CFR and is
  dependent on the bandwidth of the measurement.} The FDC and FEC are sensitive to the bandwidth or data
  acquisition rate of the
  measurement (Figure \ref{fig6}b). Whereas $W$ is insensitive to the bandwidth $W'$ is not (see Figure \ref{fig7}a). This difference is very important
  because it implies that the bandwidth dependence implicit in the
  boundary term in Eq.(\ref{eqWW'}) (the power spectrum of the force
  depends on the bandwidth if this is smaller than the corner
  frequency of the bead) is fully contained in
  $W'$. Operationally it is much easier to use $W$ rather than
  $W'$. As shown in Figure \ref{fig7}b, $W$ satisfies the CFR whereas $W'$ does
  not. The logarithm of the ratio $\log(P_F(W')/P_R(-W'))$ plotted
  versus $W'/k_{\rm B}T$ is strongly
  bandwidth dependent and exhibits a slope 30 times smaller than 1 (i.e. the slope
  expected for $W$ from the CFR) \cite{MosLorForHugRit09}.

\item{\bf How big is the error committed in recovering free
energy differences by using $W'$ rather than $W$?} Despite that $W$ and
$W'$ only differ by a boundary term (cf. Eq.\ref{eqWW'}) one can show
  that, for the case of the mechanical folding/unfolding of the hairpin, the error in
recovering free energy differences using the Jarzynski equality can be
as large as 100\% \cite{MosLorForHugRit09}. The error or discrepancy increases with the
bandwidth. Interestingly enough, for small enough bandwidths (but always
larger than the coexistence kinetic rates between the folded and unfolded
states, otherwise the folding/unfolding transitions are smeared out) fluctuations in the boundary term in Eq.(\ref{eqWW'}) are
negligible and both $W$ and $W'$ are equally good. This explains why previous
experimental tests of the CFR that used $W'$ instead of $W$ produced
satisfactory results (e.g. \cite{ColRitJarSmiTinBus05}).

\item{\bf Inequivalence between moving the trap and the pipette or chamber.} The
  non-invariance of the CFR under Galilean transformations suggests that moving the optical
  trap inside the fluidic chamber should not be necessarily equivalent to moving
  the pipette glued to the fluidic chamber. We have to
  distinguish two cases depending on whether the fluid inside the chamber is 
  dragged ({\it stick} conditions) or not ({\it slip} conditions) by the moving chamber. The two scenarios
  are physically different because in the former case the bead in the
  trap is subject to an additional Stoke force due to the motion of the
  fluid. If the fluid is not dragged by the moving chamber ({\it slip} conditions) then $y$ is the
  right configurational variable. In this case,
  $U(y,\lambda)=U_m(\lambda-y)+(k/2)(y)^2$ and the generalized
  force is equal to $F= U_m'(\lambda-y)$. Note that this $F$ is not equal
  to the instantaneous force measured by the optical trap but the
  instantaneous force
  acting on the molecule. Even in case of mechanical equilibrium the
  difference between the two instantaneous forces,
  $U_m'(\lambda-y)$ and $ky$, produces a net non-negligible difference term. If the fluid does move with the chamber ({\it stick} conditions) then $x$ is again the
  right configurational variable and we recover the main results of this
  section. Interestingly, all experiments done until now that use
  motorized stages to move chambers operate in {\it stick} conditions so we do
  not expect experimental discrepancies for the definition of the work.

\item{\bf Other cases where the work definition matters.} As we showed
  in Figure \ref{fig7} the CFR and
  the right definition of work $W$ are both amenable to experimental
  test. Another interesting example where the boundary term is relevant is
  when the force $f$ (rather than the trap position) is controlled. As
  we saw in Section~\ref{controlwork} the work in that case is given by
  $W_{X_0}=-\int_{f_i}^{f_f} Xdf$ where $X=y+x+X_0$ is the absolute trap-pipette
  distance. Because $X_0$ stands for an arbitrary origin, the work $W_{X_0}$ is also a
  quantity that depends on $X_0$. This may seem unphysical but it is
  not \cite{Pel08}. The CFR is invariant respect to the value of $X_0$ as it can be
  easily checked by writing, $W_{X_0}=W_{X_0=0}-X_0(f_f-f_i)$, and using Eq.(\ref{eqcrooks})
  gives $\Delta G_{X_0}=\Delta G_{X_0=0}-fX_0$. If the force $f$ is
  controlled, then other work related quantities such as
  $W'=\int_{x_i}^{x_f}fdx$ or $W''=\int_{X_i}^{X_f}fdX$ differ from $W$
  by finite boundary terms. Again these terms make the CFR not to be satisfied for $W'$
  and $W''$. These predictions are amenable to experimental test in
  magnetic tweezers (where the force is naturally controlled) or in
  optical tweezers operating in the force clamp mode with infinite
  bandwidth \cite{GreWooAbbBlo05} (and possibly in a
  force feedback mode with finite bandwidth as well).
\end{enumerate}

\section{A generalized fluctuation relation}
\label{gfr}
The CFR can be generalized to cases where the system is initially in
partial, rather than global, equilibrium both in the forward and the reverse protocol \cite{JunMosManRit09}. Suppose we take a system at
fixed control parameter $\lambda$ in thermal equilibrium with a bath
at temperature $T$. The probability distribution over configurational
variables $x$ is Gibbsian over the whole phase space $S$ meaning that:
$P^{\rm eq}_{\lambda}(x)=\exp(-E_{\lambda}(x)/k_{\rm B}T)/Z_{\lambda}$ with
$Z_{\lambda}$ the partition function $Z_{\lambda}=\sum_{x\in
  S}\exp(-E_{\lambda}(x)/k_{\rm B}T)$, where $E_{\lambda}(x)$ is the energy
function of the system for a given pair $\lambda,x$. We refer to this
condition as global thermodynamic equilibrium. However we might
consider a case where the initial state is Gibbsian but restricted
over a subset of configurations $S'\subseteq S$. We refer to this case
as partial thermodynamic equilibrium. Partially equilibrated states
satisfy $P^{\rm eq}_{\lambda,S'}(x)=P^{\rm
  eq}_{\lambda}(x)\chi_{S'}(x)Z_{\lambda}/Z_{\lambda,S'}$, where
$\chi_{S'}$ is the characteristic function defined over the subset
$S'\subseteq S$ [$\chi_{S'}=1$ if $x\in S'$ and zero otherwise], and
$Z_{\lambda,S'}$ is the partition function restricted to the subset
$S'$, i.e. $Z_{\lambda,S'}=\sum_{x\in
  S'}\exp(-E_{\lambda}(x)/k_{\rm B}T)$. The partial free energy is then
given by $G_{\lambda,S'}=-k_{\rm B}T\log Z_{\lambda,S'}$. Let us suppose
again the scenario depicted in Figure \ref{fig3}. Along the forward process the
system is initially in partial equilibrium in $S_0$ at
$\lambda_0$. Along the reverse process the system is initially in
partial equilibrium in $S_1$ at $\lambda_1$. The generalized CFR
reads,
\be
\frac{p_\mathrm{F}^{S_0\to S_1}}{p_\mathrm{R}^{S_0\gets S_1}}\frac{P^{S_0\to S_1}(W)}{P^{S_0\gets S_1}(-W)}=
\exp\left[\frac{W-\Delta G_{S_0,\lambda_0}^{S_1,\lambda_1}}{k_{\rm B}T}\right] \,,
\label{eqgcfr}
\ee
where the direction of the arrow distinguishes forward from
reverse, $p_\mathrm{F}^{S_0\to S_1}$ ($p_\mathrm{R}^{S_0\gets S_1}$) stands for the probability to be in $S_1$ ($S_0$) at the end of the
forward (reverse) process, and $\Delta G_{S_0,\lambda_0}^{S_1,\lambda_1}=G_{S_1}(\lambda_1)-G_{S_0}(\lambda_0)$
is the free energy difference between the partially equilibrated states
$S_0$ and $S_1$. 

\begin{figure}
\includegraphics[scale=0.4]{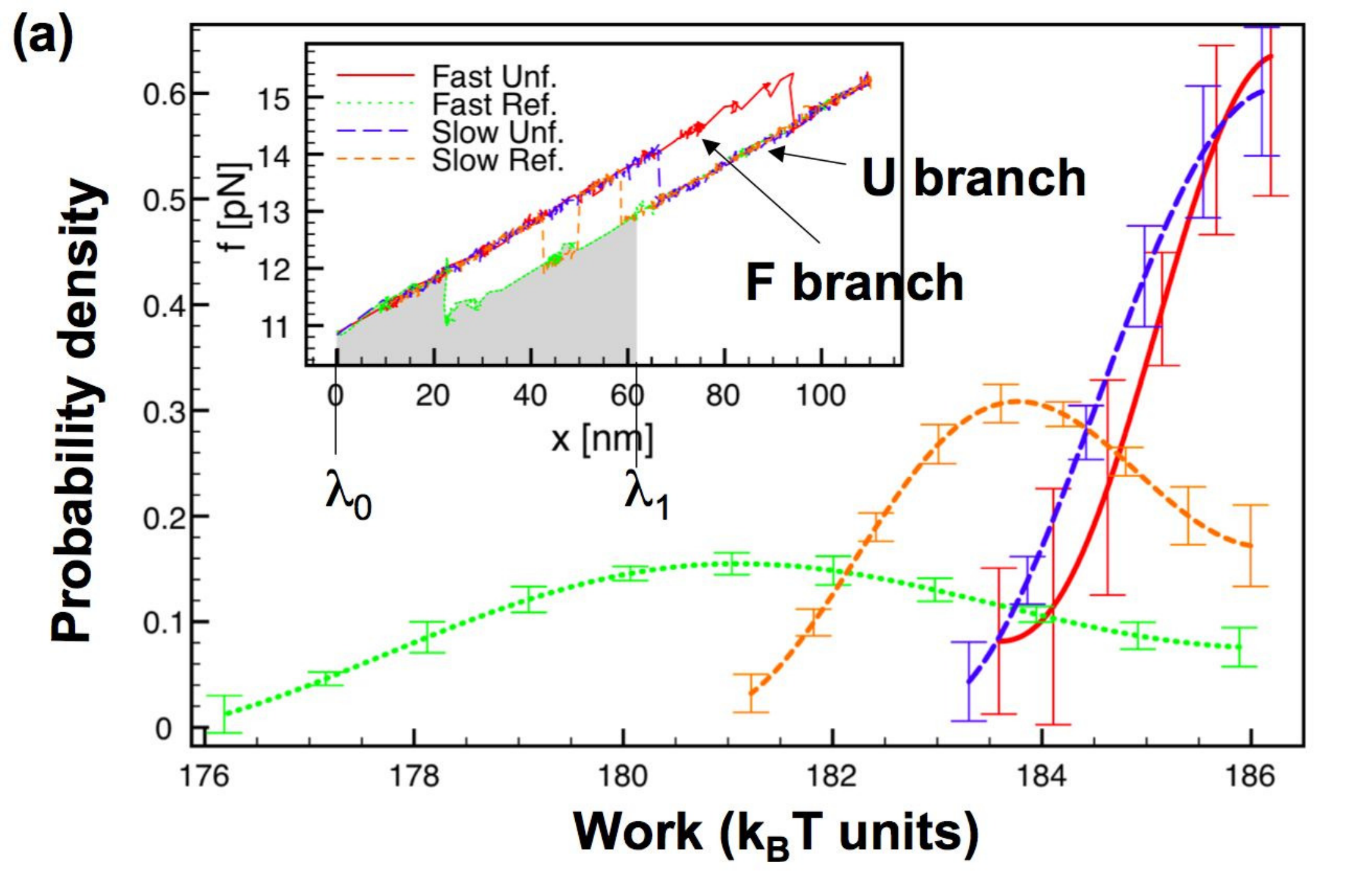}\includegraphics[scale=0.5]{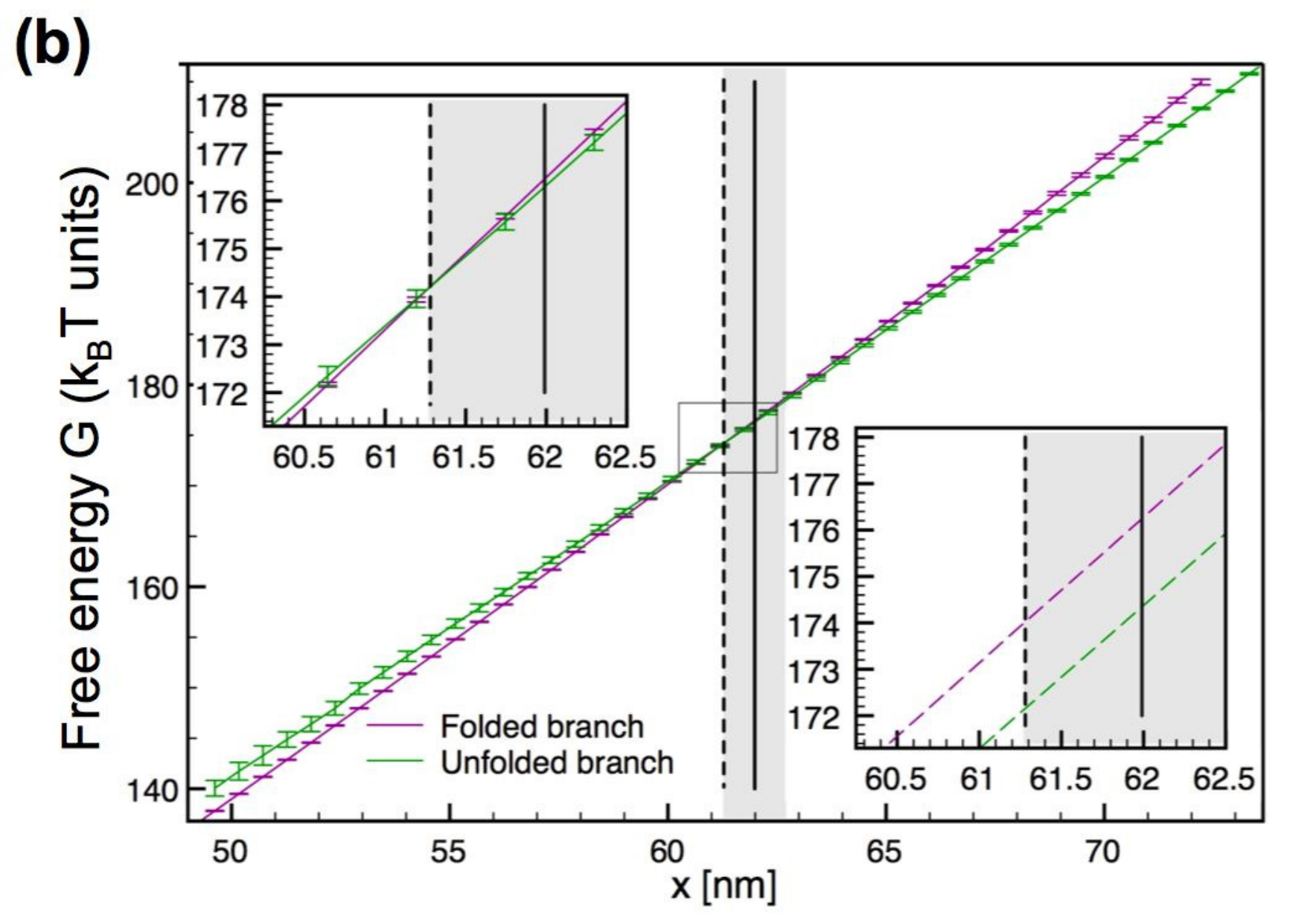}
\caption{{\bf The generalized Crooks fluctuation relation.} (a)
  Constrained work distributions measured in a 20bp hairpin at two
  different pulling speeds: 300nm/s (red, unfolding; green, refolding)
  and 40nm/s (blue, unfolding; orange, refolding). (a,Inset) The forward
  trajectories we consider are those where the hairpin starts
  partially equilibrated in the folded (F) state at $\lambda_0$ and
  ends in the unfolded (U) state (partially equilibrated or not) at
  $\lambda_1$.Note that, due to the correction term
  $p_\mathrm{F}^{F\to U}/p_\mathrm{R}^{F\gets U}$
  appearing in Eq.(\ref{eqgcfr}), restricted unfolding and refolding
  work distributions should not cross each other at a work value
  that is independent of the pulling speed. (b) Reconstruction of the
  folded (cyan color) and unfolded (green) free energy branches by
  applying the generalized CFR, Eq.(\ref{eqgcfr}), as shown in (a) and
  by varying the parameter $x\equiv \lambda_1$. The two branches cross each
  other around $x_c\simeq 82nm$ corresponding to the coexistence
  transition. For $x<x_c$ ($x>x_c$) the F (U)
  state is the minimum free energy state. The upper left inset shows
  an enlarged view of the crossing region. The lower left inset shows
  the importance of the correction term $p_\mathrm{F}^{F\to
      U}/p_\mathrm{R}^{F\gets U}$ appearing in
  Eq.(\ref{eqgcfr}). If that term was not included in
  Eq.(\ref{eqgcfr}) the coexistence transition disappears. Figure taken from \cite{JunMosManRit09}.}
\label{fig8}
\end{figure}

Partially equilibrated states appear in many cases, from thermodynamic
branches to  intermediate and
  misfolded molecular states . The usefulness
  of the generalized CFR relies on our possibility to experimentally distinguish the
  substates visited along any trajectory and that these substates be visited frequently enough. For example, a
  molecule pulled by stretching forces can be in partial equilibrium
  when it stays either in the
  folded or unfolded state until it transits to the other state. If
  $S_0$ stands for the folded state and $S_1$ for the
  unfolded state, the generalized CFR makes possible to extract
  the free energies $G_{S'}(\lambda)$ of the folded and
  unfolded states $S'=S_0,S_1$ along
  the $\lambda$-axis, i.e. the folded and unfolded
  branches. Figure \ref{fig8} shows an experimental verification of this
  result for a DNA haiprin that folds/unfolds in a two-states manner.

\section{Conclusion}
The possibility to experimentally measure the inverted motions remarked
by Schr\"odinger more than half a century ago has boosted the study of
energy fluctuations in very small objects under nonequilibrium
conditions. The possibility to measure work fluctuations in single
molecules that are mechanically unfolded has provided the testing ground
for some of the most recent theoretical developments in nonequilibrium
statistical physics. Fluctuation relations and fluctuation theorems
(e.g. the Gallavotti-Cohen theorem for steady state systems \cite{GalCoh95}) are examples of new results that
quantify {\it inverted motions} in nonequilibrium
states. Measuring {\it inverted motions} has also practical applications: the
Crooks fluctuation relation (CFR), Eq.(\ref{eqcrooks}), and its
generalization to partial (rather than full) equilibrium conditions, Eq.(\ref{eqgcfr}),
allows us to extract free energy differences between native or
non-native states and free energies of thermodynamic branches. Future
studies will also show the reliability of these methodologies to extract
free energies of misfolded and intermediate states in RNAs or proteins,
and base-pair free energies in nucleic acids unzipped under irreversible
conditions. We also stressed how important is the correct definition of
mechanical work to ensure the validity of the CFR. In this regard
serious inconsistencies are encountered using other definitions of
mechanical work but such inconsistencies are
nowadays amenable to experimental test. Related to this, it is also important to
underline the general non-invariance of fluctuation relations and
theorems under Galilean transformations \cite{SpeMehSei08}, an aspect that has not been
stressed enough and that can also be
tested in experiments. Finally, all the studies covered in this note
address energy fluctuations of small classical systems under Gaussian
noise conditions. It would be highly desiderable to have experiments
done in systems in the regime of non-Gaussian noise (maybe at
submicroseconds timescales), or in quantum systems where the concept of
classical trajectory looses its usual meaning \cite{Ritort09}.

\begin{theacknowledgments}

A. A. and M. R. acknowledge financial support from the Spanish MEC-MICINN through the FPU fellowhip program, grant no. 
AP2007-00995 and Human Frontier Science
Program (HFSP) (RGP55-2008) respectively. F. R acknowledges financial support from
Grants FIS2007-3454. Human Frontier Science Program (HFSP)
(RGP55-2008) and Icrea Academia prize 2008. 
\end{theacknowledgments}

\bibliographystyle{aipproc}   % if natbib is available

\end{document}